\documentclass[preprint,12pt]{elsarticle}
\usepackage{graphicx}
\usepackage{amssymb}
\usepackage{lineno}

\journal{Chinese Physics C}

\begin{document}

\begin{frontmatter}

\title{The silicon matrix for the prototype of the Dark Matter Particle Explorer}

\author{R.R.Fan, F.Zhang, W.X.Peng, Y.F.Dong, K.Gong, S.Yang, D.Y.Guo, J.Z.Wang, M.Gao, X.H.Liang, J.Y.Zhang, X.Z.Cui, Y.Q.Liu, H.Y.Wang}

\address{Institute of High Energy Physics Chinese Academy of Sciences, Beijing P.R.China}

\begin{abstract}
A new generation detector for the high
energy cosmic ray
- the DAMPE(DArk Matter Particle Explorer)
is a satellite based
project. Its main object is the energy
measurement of the cosmic ray
nuclei from 100GeV to
100TeV, the high energy electrons and  gamma
ray from 5GeV to 10TeV.
A silicon matrix detector
described in this paper, is
employed for the sea level cosmic ray energy and position
detection while the prototype testing of the DAMPE.
It is composed by the 180 silicon PIN detectors,
which covers an area of ${32*20 cm^{2}}$.
The primary
test result shows
a good MIPs discrimination.
\end{abstract}


\end{frontmatter}

\linenumbers

\section{Introduction}
\label{Introduction}
With the results of AMS-02\cite{1}, ATIC\cite{2}, PAMELA\cite{3}
and FERMI\cite{4}, more and more
proofs pointed the electron excess of cosmic
electrons. With a requirement
of higher and more precise
energy measurement, a dark matter search satellite
project DAMPE has been
put forward.

The main scientific objective of DAMPE is to measure
electrons and photons with much higher energy resolution
and energy reach than achievable with existing space
experiments in order to identify possible Dark Matter
signatures. It has also great potential in advancing
the understanding of the origin and propagation mechanism
of high energy cosmic rays, as well as in new discoveries
in high energy gamma astronomy.

\begin{figure}
  \includegraphics[width=5in]{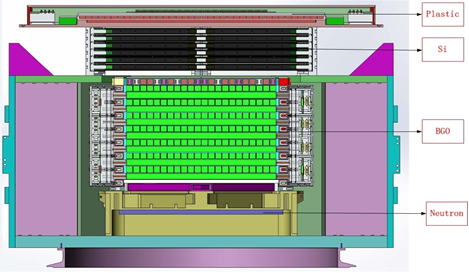}
  \caption{The schematic of DAMPE detector: The components of the DAMPE are PSD, STK, BGO and
the neutron detector from top to
bottom.}\label{DAMPE.fig}
\end{figure}

There are four sub-detector in the DAMPE system, the plastic scintillator(PSD),
the silicon tracker(STK), the BGO electromagnetic calorimeter(BGO) and
the neutron detector.(see Figure \ref{DAMPE.fig})

The plastic scintillator and
silicon tracker are employed to
charge particles'
discrimination and
reconstruction the track. The
silicon tracker with tungsten
convertors can also
distinguish the gamma ray and
electrons. The STK is followed by an imaging calorimeter
of about 31 radiation lengths thickness, made up of 14 layers
of BGO bars in a hodoscopic arrangement. A layer of neutron
detectors is added to the bottom of the calorimeter for the proton/electron discrimination.
The total thickness of the BGO calorimeter and the STK
correspond to about 33 radiation lengths, making it the
deepest calorimeter ever used in space.

For electrons and photons, the detection range of the DAMPE is
5 GeV - 10 TeV, with an energy resolution of
about 1\% at 800 GeV. For cosmic rays, the
detection range is 100 GeV - 100 TeV,
with an energy resolution better than 40\%
at 800 GeV. The geometrical factor is about
${0.3 m^{2}sr}$ for electrons and photons,
and about ${0.2m^{2}sr}$ for cosmic rays.
The angular resolution is 0.1$^\circ$ at 100 GeV.

During the DAMPE prototype testing,
a silicon matrix built by the
Institute of High Energy
Phsics(IHEP), the Chinese Academy of Sciences(CAS),
serves as the silicon tracker.
Followed chapters describe the details of the silicon
matrix construction and
test.
\section{The detector structure}
\label{matrix}

\subsection{the silicon matrix}
\begin{figure}
  \includegraphics[width=5in]{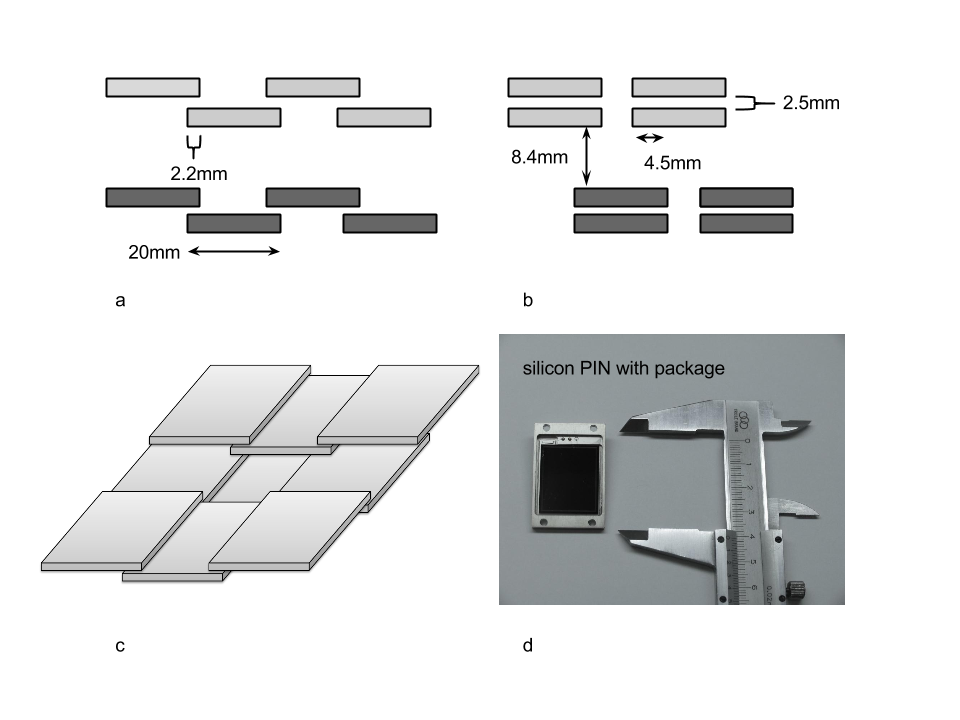}
  \caption{Illustrations of the matrix structure:
  a,b,c are different axes view of the silicon matrix.
  d is the photo of the silicon PIN detector with package.}\label{det.fig}
\end{figure}

In accordance of the required sensitive area, the
matrix is composed of 180
large area
silicon PIN detectors(see Figure \ref{det.fig} d).
Each of them
has an active area of ${ 20*25mm^{2}}$ and 0.5mm thickness.
The full depletion voltage
for the selected detectors is 40 V, while the
operation voltage is 60 V.
The measure leak current of detectors is under
100 nA.
The detector after selection
is attached on a ceramic chip in a Kovar package.

To achieve a better detection efficiency of incident particles over $\pm$60$^\circ$ ,
every detector has a small overlap to the neighbor ones.(see Figure \ref{det.fig} c)
The overlap is determined by the calculation and confirmed by the fast
simulation.

\begin{figure}
  \includegraphics[width=5in]{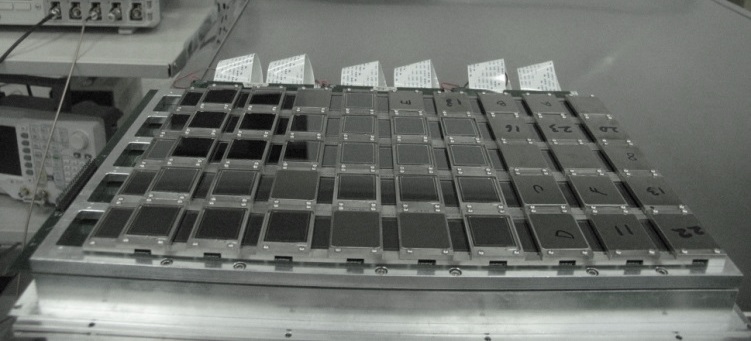}
  \caption{The photo of the top layer of the matrix: the 90(18*5) silicon PINs on top of the matrix,
   which is above the bottom layers.}\label{matrix.fig}
\end{figure}

While the assembling, the
detector is fixed on the
readout PCB and an aluminum
frame via four screws.
The detectors on top and on bottom are combined
in two layers respectively.
Figure
\ref{matrix.fig} shows the
photograph of the top layer matrix.
As this structure, the silicon
matrix can achieved an active area up to ${32*20 cm^{2}}$.

\subsection{the electronics}
\begin{figure}
  \includegraphics[width=5in]{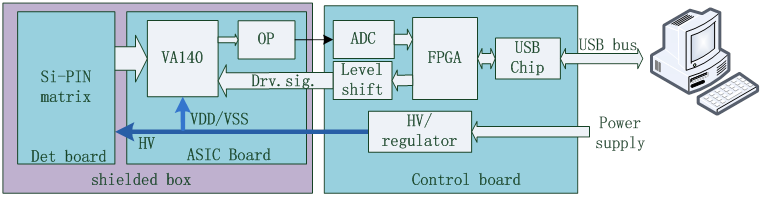}
  \caption{The readout schematic of the matrix: the matrix and the FEE(ASIC board)
  are placed in a shield box, the control board transfer the data to PC via USB bus.}\label{electronics.fig}
\end{figure}

As shown in Figure \ref{electronics.fig},
the readout electronics system of the matrix
consists an ASIC PCB and a control PCB.
A aluminum box is employed to shield the detectors
and the front-end electronics(FEE) from the surrounding
light and the electro-magnetic interference.
A VA140 readout PCB is
mounted on the back of each
layer.
The signals through the flex are read out by six VA140 chips,
which are controlled by an FPGA.
The DAQ communicated with the computer via the USB bus.

The VA140 is a 64-channel,
low noise and high dynamic range
charge measurement ASIC designed
by IDEADS (Norway)\cite{5}. It is an updated version
of the VA64HDR9A ASIC, which is used by AMS\cite{6},
implemented in 0.35um process
for lower power consumption and better radiation tolerance.
Each channel contains a charge sensitive preamplifier,
a shaper circuit and a sample-hold circuit.
Since the primary package can
only afford 32 pins output,
we used the half of all the 64 channels to
readout signals in every VA140.

The control board is used
for the VA140 control, the analog signal
digitalization
and the data communication. It contains a power
module (the HV and the regulator), a controlling FPGA(XC3S500E),
a 14-bit 3MSPS ADC(AD9243) and a USB interface chip(CY7C68013).
A level shift circuit from $0\sim~3.3V$ to $-2V\sim+1.5V$
is used in the control board,
due to the mismatched level voltage
between the FPGA and the VA140.

The DAQ is based on a LabWindows GUI
online software, while the
offline data analysis is undertaken
with the ROOT framework.
In the test of electronics,
An 100Hz random trigger is
induced in to acquire the
pedestals of every channel.
We get an average noise of
the channels is about 0.35 fC.

\section{The cosmic ray test}
\label{Results}

\begin{figure}
  \includegraphics[width=5in]{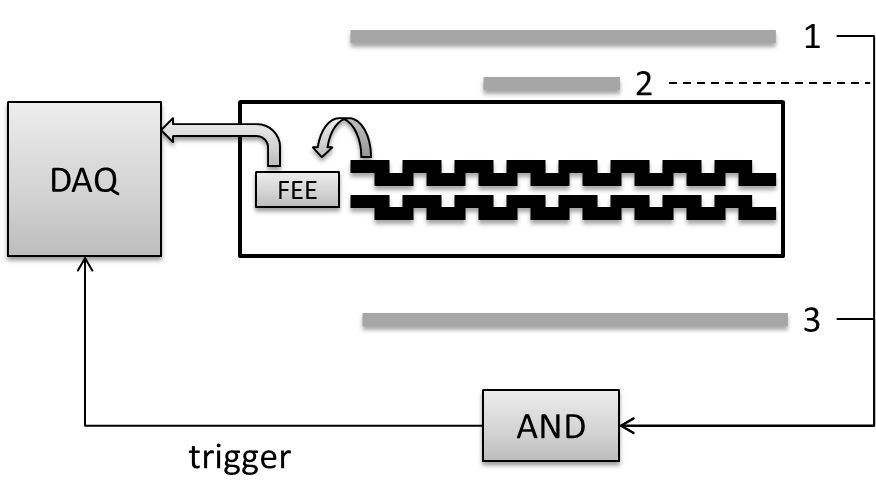}
  \caption{The telescope system of the cosmic ray test:
  the 1,2,3 stand for three scintillators, which give the trigger to the DAQ.
  The matrix is in the middle of these scintillators.
  }\label{trigger.fig}
\end{figure}

There are two requirements of
this matrix, giving the
particle charge information
and the track reconstruction.
The test in
laboratory is completed with
the sea level cosmic ray.
In this test, we can get the
unit charge response(MIPs)
and an approximately position
distribution.

A simple telescope
system is set up for the
test, which can be seen in
Figure \ref{trigger.fig}. The
grey box
labeled 1,2,3 stand for plastic
scintillators. The area of scintillator 1 and 3 is
${40*40cm^{2}}$, which covers the whole surface of the matrix.
The silicon matrix is placed
in the middle of the plastic
scintillators. When the
incident particle pass through,
this two plastic scintillators
give the coincidence trigger to the DAQ.
With a long time test, the MIPs energy spectrum can be obtained in readout channels.
To get a position sensitive plot, a smaller scintillator 2 with a area of
${12*10cm^{2}}$ is inserted into the scintillator 1 and 3 as the indicator.

\begin{figure}
  \includegraphics[width=5in]{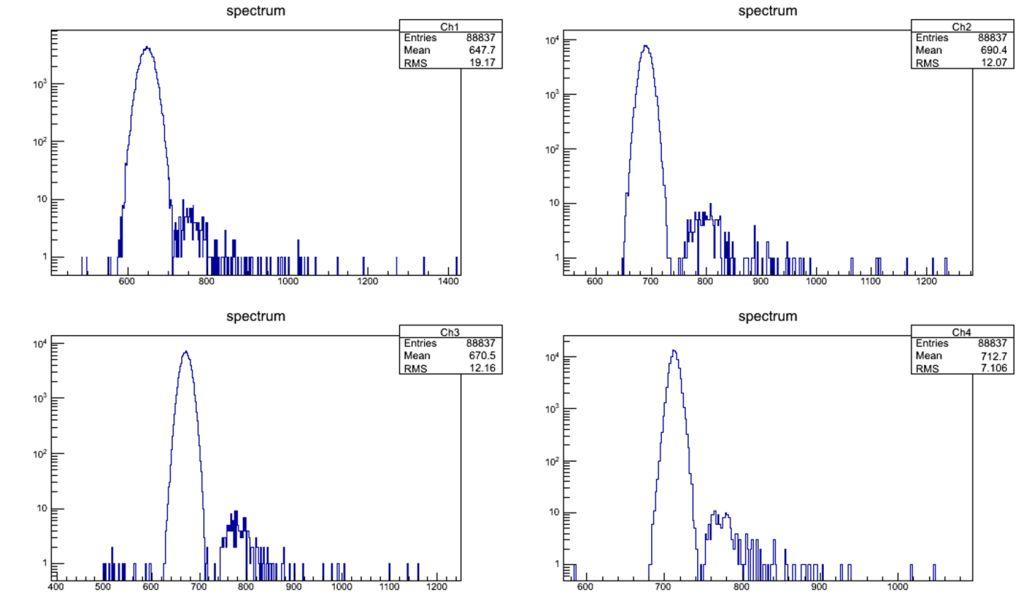}
  \caption{The spectrums of MIPs in the first four channels:
  the X axis is ADC number, the Y is counts.
  }\label{spectrum.fig}
\end{figure}

\begin{figure}
  \includegraphics[width=5in]{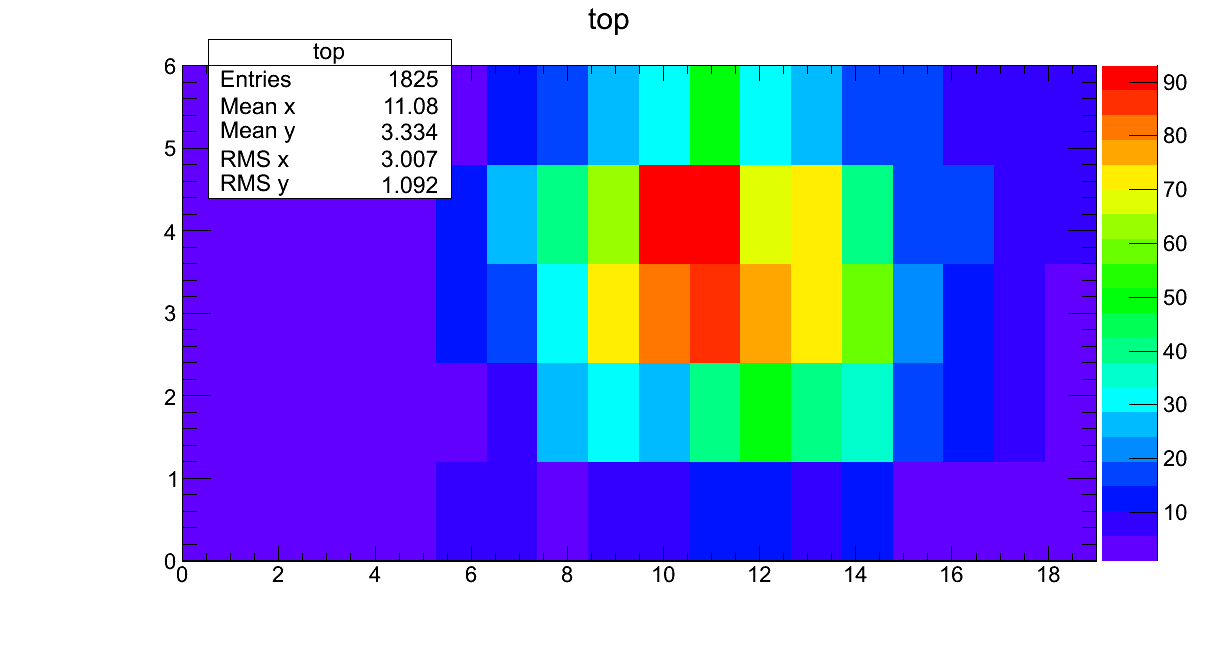}
  \includegraphics[width=5in]{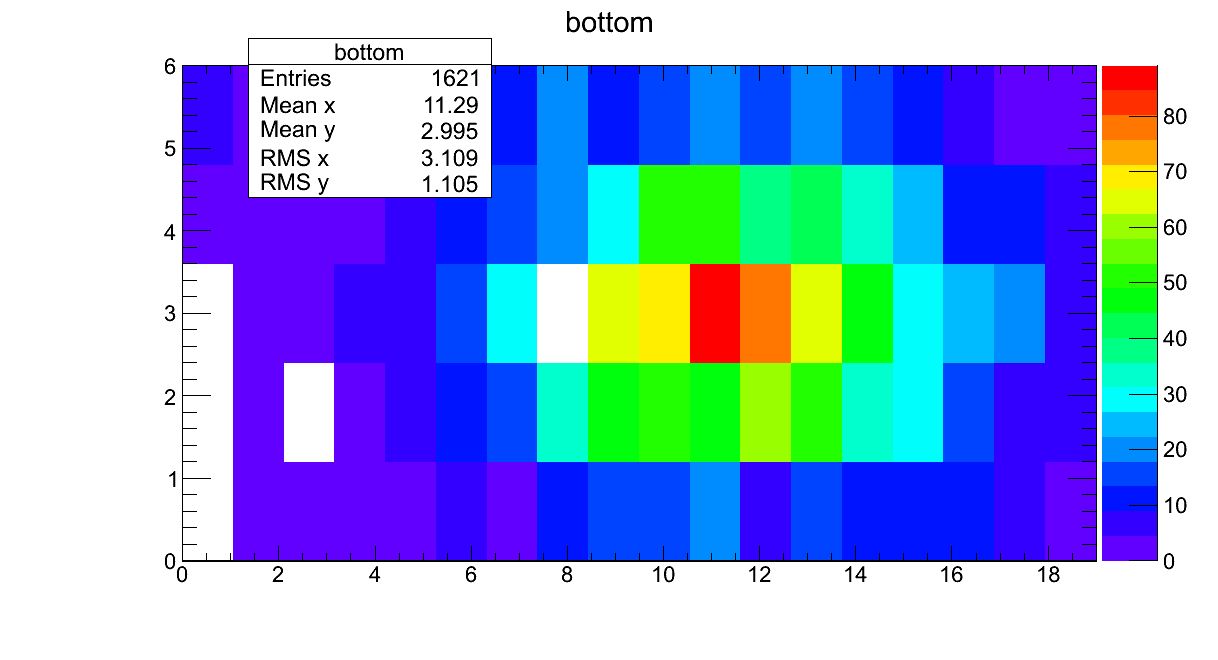}
  \caption{The lego plot of the position sensitive test: the X and Y axis unit is the
detector channel. The palette gives the counts of MIPs}\label{cosmic.fig}
\end{figure}

After 10 hours
cumulating, the spectrum of
each channel is obvious.
Because of the angle's
effect, the charge resolution
can't be directly derived from these
spectrums. We still get
clear views of the MIPs' peak in 175
channels(97\%).
Figure \ref{spectrum.fig} shows the first
4 channels' cumulate spectrums.

In the position sensitive
testing, we got the MIPs
counts map both of the top and
the bottom layer(see Figure \ref{cosmic.fig}).
In the plot, the X and Y axis unit is the
readout channel. The
color shows different counts
of the MIPs in each detector.
In the figure, we can see that
except 5 bad channels in
the bottom layer(the white pixels),
the gravity of two
layers is almost same in X
axis. An offset about half of Y unit
exists in the Y axis, because of the
detector structure(see Figure \ref{det.fig}). The spot covers an
area about 5*3, which is in accord with
the small-sized scintillator.

\section{Conclusion}
\label{Conclusion}
With the
cosmic ray test, the matrix
can distinguish the MIPs
signal from the electronics
noise clearly, while the
position can be confirmed by
the channel sequence.
Through a long time running, the
matrix is optimized for the
DAMPE prototype. The
results of the DAMPE prototype
test will be presented in near
future.

\section{Acknowledgements}
\label{Acknowledgements}
This work is supported by the Chinese Strategic Priority
Research Program in Space Science, CAS.
The authors also greatly
appreciated the collaboration with the DAMPE colleagues
from the University of Science and
Technology of China(USTC) and
the Purple Mountain
Observatory(PMO) during the electronics and the DAMPE prototype test.





\bibliographystyle{model1-num-names}
\bibliography{<your-bib-database>}



\end{document}